\documentclass[a4paper,fleqn]{cas-dc}
\usepackage{nccmath}
\usepackage{amsmath}
\usepackage[numbers]{natbib}
\usepackage{dcolumn}
\usepackage{mathtools}
\usepackage{bm}

\begin{document}
\let\WriteBookmarks\relax
\def\floatpagepagefraction{1}
\def\textpagefraction{.001}

 \shorttitle{Reminiscences about Hans Capel}


\title[mode=title]{Reminiscences about Hans Capel}   




%



\ead{tsallis@cbpf.br}




\affiliation[1]{organization={Centro Brasileiro de Pesquisas Fisicas and National Institute of Science and Technology for Complex Systems}, 
    addressline={Rua Xavier Sigaud 150}, 
    city={Rio de Janeiro}, 
    postcode={22290-180}, 
    country={Brazil}}

\affiliation[2]{organization={Santa Fe Institute}, 
    addressline={1399 Hyde Park Road}, 
    city={Santa Fe}, 
    postcode={87501}, 
    country={USA}}

\affiliation[3]{organization={Complexity Science Hub Vienna}, 
    addressline={Metternichgasse 8}, 
    postcode={1030}, 
    state={Vienna}, 
    country={Austria}}

\affiliation[4]{organization={Dipartimento di Fisica e Astronomia Ettore Majorana, University of Catania}, 
    addressline={Via S. Sofia 64}, 
    postcode={95123}, 
    state={Catania CT}, 
    country={Italy}}

\author[1, 2, 3, 4]{Constantino Tsallis}[]






\begin{abstract}
As an homage to the memory of Hans Willem Capel, I present some personal reminiscences and thoughts that come to my mind in this special occasion.
\end{abstract}

\begin{keywords}
Hans Willem Capel; 
Blume-Capel model; 
Tricritical points;
Physica A Editors
\end{keywords}                          
\maketitle

\section*{Remembering Hans}
I believe that I first met Hans Willem Capel around 1986, in the context of the Statphys meeting held in Boston, when he invited me to become a member of the Editorial Board of Physica A.  Since then, I met him many times in Amsterdam and elsewhere. He eventually invited me to become, together with Gene Stanley and Giorgio Parisi,  Main Editor of Physica A, which I gladly accepted, having then various opportunities to directly interact with all of them and various other distinguished colleagues, including Kenneth A. Dawson and 
the present Editor-in Chief Joseph Indekeu, whom we collectively invited to join us in the editorial tasks.
Hans and I eventually became friends and the Physics Department of the University of Amsterdam gradually became for me a privileged place where we shared and implemented all kinds of editorial activities and decisions.

Along the years, we naturally had several scientific conversations and, more than once,  he invited me (1987, 1998, \dots) to deliver seminars at his working place. Indeed, phase transitions and critical phenomena since long was -- and still is -- a most interesting subject for exchanging all kinds of physical insights. However, quite regretfully, it never happened that we co-author a paper. In any case, my interest for the well known Blume-Capel model \cite{Blume1966,Capel1966a,Capel1966b,Capel1967} was very neat at the time and I even seriously analyzed, for some time,  the possibility of approaching it through renormalization-group techniques. For whatever reason, it did not happen that
our specific scientific research projects overlap, which I would have certainly appreciated. Since 1988, when I started focusing on grounding statistical mechanics on nonadditive entropic functionals \cite{Tsallis1988,Tsallis2009,Tsallis2023}, I gradually diminished my activities in the traditional theory of phase transitions. This was obviously not particularly favorable for initiating a possible collaboration with Hans. However, very recently, I did notice with pleasure that an overlapping of our research areas did emerge spontaneously, undertaken by colleagues from India \cite{overlap}. Indeed, they discussed the validity and possible failure of the Boltzmann-Gibbs (BG) theory for the Blume-Capel model along some specific physical regimes.  

In one occasion, in USA, I coincidentally met Martin Blume, Editor-in-Chief of the American Physical Society's Physical Review journals at the time. We occasionally talked about Hans and I was delighted to see the high opinion that Blume had about Capel.

Hans had in fact a very friendly personality, always ready to be helpful in any kind of matter, always welcoming with a smile all sorts of suggestions or requests.


\section*{A thought about the deep meaning of entropy}
We all know people whose working office appears to be an amazing mess. Hans definitive was one of those, I was always impressed by the various tall columns of letters and manuscripts editorially submitted to Physica A that were displayed on his table and elsewhere. In spite of the feeling of great disorder that this view generated to observers like me, Hans knew apparently how to quickly find any specific manuscript or information. This fact makes me think about one of the -- occasionally controversial -- aspects essentially associated with the concept of entropy. If I was hypothetically asked to evaluate the entropy of Hans's office, I would attribute to it a very high value, in strong discrepancy with the evaluation that Hans would have provided to the same question, which most probably would yield a quite low value. This interesting property directly reflects the fact that information about a given system typically is different for different observers and different circumstances. This fact is consistent with the famous claim by Edwin T. Jaynes, along Eugene P. Wigner lines, that entropy is an {\it anthropomorphic concept}. In 1965, he wrote \cite{Jaynes1965}:

{\it Entropy is an anthropomorphic concept, not only in the well-known statistical sense that it measures the extent of human ignorance as to the microstate. Even at the purely phenomenological level, entropy is an anthropomorphic concept. For it is a property, not of the physical system, but of the particular experiments you or I choose to perform on it.}

At this point, it is relevant to emphatically distinguish {\it entropic functional} from {\it entropy}. 

The former (i.e., entropic functional) merely is a function of a set of probabilities $\{p_i\}$, e.g., the Boltzmann-Gibbs-von Neumann-Shannon (additive) expression
\begin{equation}
S_{BG}=-k\sum_{i=1}^W p_i \ln p_i \,,
\end{equation}
or its (generically nonadditive) $q$-generalization \cite{Tsallis1988}
\begin{equation}
S_{q}=k\sum_{i=1}^W p_i \ln_q \frac{1}{p_i} \;\;\;(S_1=S_{BG})
\end{equation}
with
\begin{equation}
\ln_q z \equiv \frac{z^{1-q}-1}{1-q} \;\;\;(\ln_1 z=\ln z)\,,
\end{equation}
or its (generically nonadditive) $\delta$-generalization \cite{Tsallis2009}
\begin{equation}
S_{\delta}=k\sum_{i=1}^W p_i \Bigl[\ln \frac{1}{p_i} \Bigr]^\delta \;\;\;(S_1=S_{BG}) \,,
\end{equation} 
or even the (generically nonadditive) unification \cite{TsallisCirto2013} of all the above
\begin{equation}
S_{q,\delta}=k\sum_{i=1}^W p_i \Bigl[\ln_q \frac{1}{p_i} \Bigr]^\delta \;\;\;(S_{q,1}=S_q; S_{1,\delta}=S_\delta; S_{1,1}=S_{BG}) \,.
\end{equation}

The latter (i.e., entropy) consists in expressing a specific entropic functional in terms corresponding to a specific class of theoretical models, or a specific class of experiments (with a given precision), or a specific scale of description, such as a traditional microscopic Hamiltonian approach (say that of a condensed matter theoretician), or, along a very opposite description, namely at the thermodynamic, macroscopic scale (say that of an engineer), or even at the various ensemble choices, namely the microcanonical (equal probabilities of the admissible states, which corresponds to an isolated physical system), canonical (physical system in contact with a thermal reservoir at given temperature $T$), and grand-canonical ones. The primary choice always is an entropic functional which, for special values of its entropic indices, implies entropic extensivity (in the thermodynamical sense) for the specific class of systems that is being focused on. 

A clear illustration of the convenience of distinguishing entropic functional from entropy is provided by the so-called Barrow entropy $S_B$ \cite{Barrow2020}. This entropy corresponds to a fractal generalization of the Bekenstein-Hawking entropy $S_{BH}$ for a black hole (microcanonical description, i.e., assuming equal probabilities), and it can be expressed as $S_B=k(S_{BH}/k)^\delta$ ($\delta$ is sometimes expressed as $\delta \equiv 1+ \Delta/2$). Barrow entropy is a plausible, phenomenological entropy heuristically introduced to satisfy astrophysical/cosmological requirements, but it is relevant to realize that it is not grounded on any entropic functional. In other words, the Barrow entropy is well defined in spite of the fact that no Barrow entropic functional exists. Let us also mention that the microcanical ensemble choice for the entropic functional $S_\delta$ (as well as for the composable entropic functional used in \cite{TsallisJensen2025}) also provides $S_\delta=k(S_{BH}/k)^\delta$, thus coinciding with $S_B$\footnote{This coincidence is at the origin of the confusing expression "Barrow-Tsallis entropy" or "Tsallis-Barrow entropy" which, regretfully, appears occasionally in the literature.}. This simple connection would most probably disappear if we were to consider an ensemble different from the microcanonical one \cite{Tsallis2026}.

To close this Section, I would like to narrate an unforgettable scene that I witnessed in April 2000 at the North Texas University, Denton TX. Michel Baranger, MIT's Emeritus Professor at the time, was delivering an interesting colloquium at the NTU Physics Department. As I remember, he said something close to "I have a many-body system whose admissible Gibbs phase-space consists of a highly-dimensional box within which I am following the precise trajectory of the system. The entropy remains strictly zero all the time, since, at all times, we know everything that it is to be known. At a certain point, you get tired of following this precise microscopic trajectory and decide to only state that it remains within the volume of that box, therefore the entropy consistently jumps to a neatly non-vanishing positive value." Then, standing up, firmly faced the audience, and rhetorically asked "Who made the entropy to increase?" Then, he terminated "You did!". That was just wonderful! Dieter H. E. Gross, a distinguished, traditional German physicist, was in the audience, and he immediately shouted something like "That is absurd, absolutely inadmissible!" I must confess -- noblesse oblige! -- that I fully share Baranger's understanding, against Gross'understanding.  
 
When I am occasionally asked what is the most subtle concept in physics, I tend to reply that the quantum wave function is a very subtle one, but that I believe that entropy is even more. I base this perception on the observation that, analyzing the concept of wave function, two good pysicists will start diverging after say five minutes of conversation. But, if they analyze the concept of entropy, they will most probably start diverging after say three minutes! The most powerful motor of science definitely is not what we know, but what we do not know!

\section*{A technical issue that I would have loved to discuss with Hans}
Close to 30 years ago, I received in Rio de Janeiro the visit of a dear colleague and friend from Oxford University, Robin B. Stinchcombe, a notorious expert in geometric and thermal phase transitions. We lengthily discussed about a possible connection between the usual theory of critical phenomena and nonadditive entropic functionals. We both agreed that such connection ought to exist  but this remained as an elusive point until very recently \cite{mariano1,mariano2,Tsallis2025Uzbek,LimaTsallis2025,Tsallis2025Chaos}. Indeed, it gradually became neat and clear that, for standard critical phenomena in $d$-dimensional short-range-interacting many-body Hamiltonians (say Ising, XY, Heisenberg models), Boltzmann-Gibbs statistical mechanics, grounded on the traditional Boltzmann-Gibbs-von Neumann-Shannon additive entropic functional $S_{BG}$ works -- as well known -- very well in both ordered and disordered phases (see, for instance, \cite{Reichl2016}), but it fails at precisely the critical point. The calculation of all the anomalous critical exponents describing the {\it approach} to the critical point ($\alpha, \beta, \gamma, \nu, \dots$) is one among the many impressive successes  of the BG theory. But, what happens at {\it precisely} the critical point, where quantities such as the specific heat, isothermal magnetic susceptibility, correlation length, Gruneisen parameter, and similar ones diverge (or vanish, like say the spontaneous order parameter\footnote{For a physical (thermodynamical)  quantity $\xi$ which takes positive real values, infinity is equally inadmissible as zero, given that each of them is the inverse of the other one. In the spirit of Karl Popper's {\it falsifiability} requirement \cite{Popper1959}, a typical  experimental result should be expressed as $\xi_{0} \pm \epsilon_{\xi}$, where $\xi_0 \in (0,\infty)$. This is not possible if $\xi_0$ appears to be infinity or zero. In the former (e.g., the thermal magnetic susceptibility at the critical point), the falsifiability requirement must be expressed as $\xi \ge  \xi_0 \gg 1$; in the latter (e.g., the spontaneous magnetization at the critical point, or the mass of neutrinos in high-energy physics), as $ \xi \le \xi_0  << 1$. In the context of Popper's requirement, such cases may be seen as borderline.})  is a different and definitively delicate issue. The critical exponent $\delta$, defined so as to characterize the behavior of the order parameter in the presence of a nonvanishing field -- the field which is thermodynamically conjugate of the order parameter itself -- {\it precisely at $T_c$}, represents a paradigmatic such example. In fact, this issue has since long started to be focused on: see \cite{Robledo1999} as well as \cite{Robledo2005}, where, for a special class of systems, a connection has been established between $\delta$ and the entropic index $q$, namely $q=\frac{1+\delta}{2}$ (hence, $\delta \ge 1$ implies $q \ge 1$).

\begin{figure}
\centering
\includegraphics[width=8.4cm]{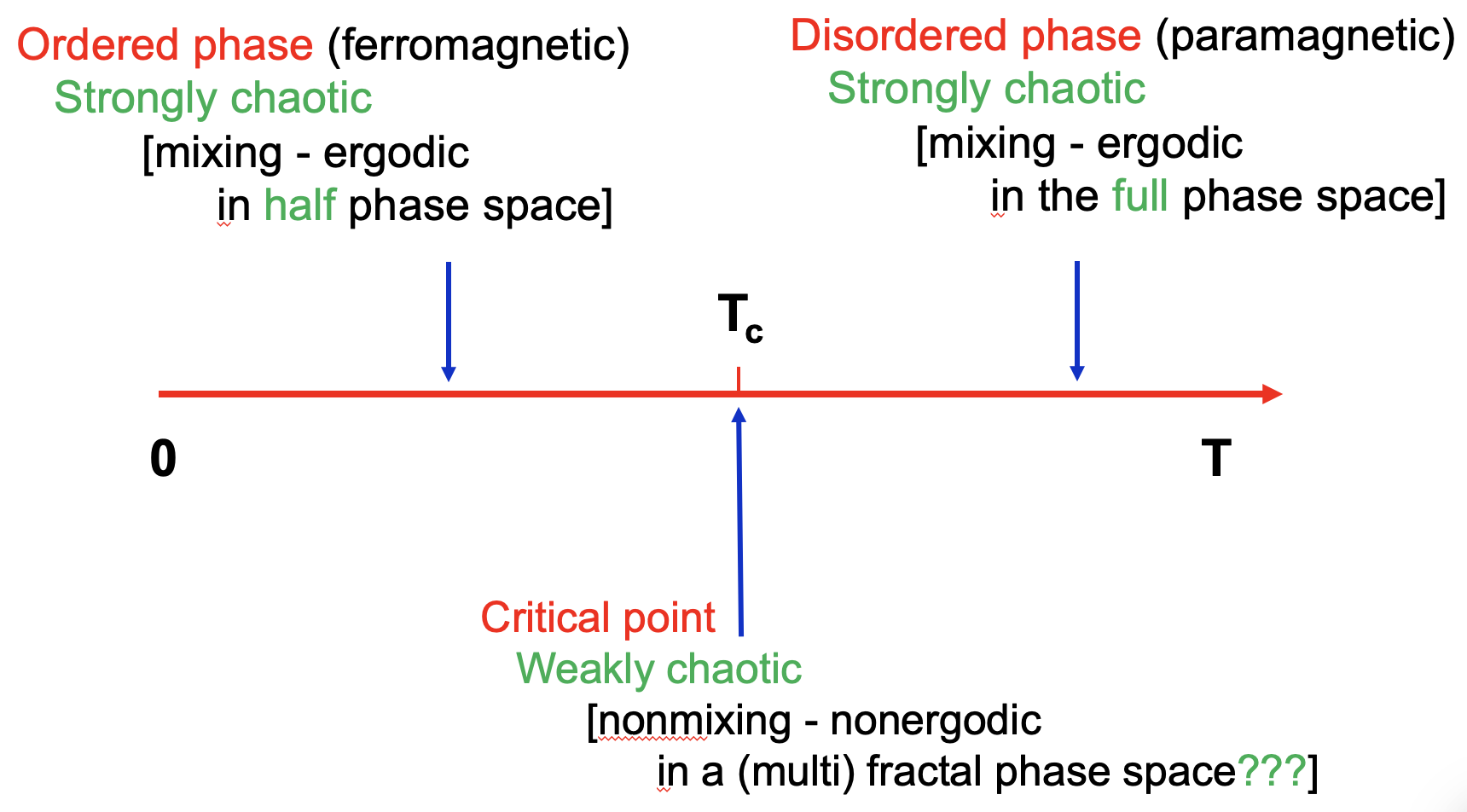}
\centering
\caption{As a paradigmatic illustration we may think of a prolate anisotropic (cigar-like, satisfying an axial symmetry) XY or Heisenberg next-nearest-coupled $d=3$ ferromagnetic model. Through the critical point ($T_c$) emerging at the $N\to\infty$ limit, the system undergoes the breakdown of the up-down symmetry corresponding to the disordered (paramagnetic) phase.}
\label{Fig1}
\end{figure}

\begin{figure}
\centering
\includegraphics[width=8.4cm]{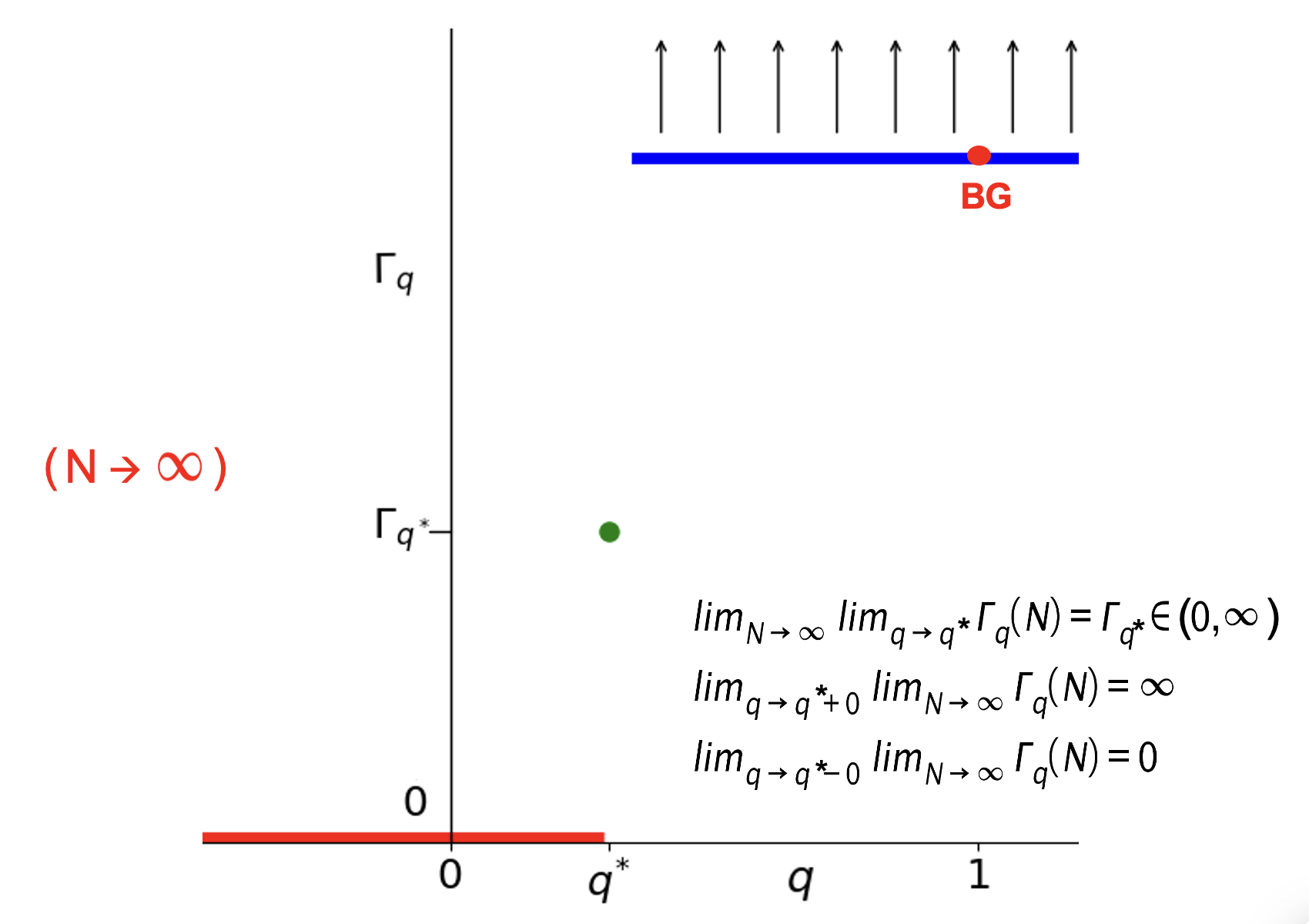}
\centering
\caption{$\Gamma_q$ diverges along the blue line ($q>q^\star$), vanishes along the red line ($q<q^\star$), and is finite on the green dot ($q=q^\star$). The red dot corresponds to the BG theory ($q=1$). When universal quantities such as the dimensionality $d$ and the symmetry which is spontaneously broken vary (e.g., $d=2,3$ and XY or Heisenberg symmetries), then the value of $q^\star$ changes. Furthermore, if different nonuniversal quantities such as spin size and neighborhhod couplings vary within the same given symmetry (say XY one), the entropic index $q^\star$ remains invariant whereas quantities such as $\Gamma_{q^\star}$ do change. }
\label{Fig2}
\end{figure}

Close to the critical point, the BG theory provides predictions which are fully confirmed by the experimental data; this by no means is surprising since ergodicity is valid under those circumstances. Indeed, in the disordered phase (e.g., the paramagnetic one, for ferromagnets) the many-body dynamics is ergodic in the entire Gibbs phase space (for classical systems); in the ordered phase (e.g., the ferromagnetic one, for ferromagnets) the system remains ergodic in a non-vanishing Lebesgue measure of the Gibbs space (half of the space if the system has an Ising-like, axial, symmetry). These features are guaranteed by the fact that, in both phases, the maximal Lyapunov exponent is positive ({\it strong chaos}), hence hypothesis like mixing and ergodicity certainly are valid). But, at precisely the critical point, the maximal Lyapunov exponent vanishes (edge of chaos,  {\it weak chaos}), and an anomalous, possibly  (multi)fractal occupancy of the phase space is expected to be the rule, corresponding to a vanishing Lebesgue measure of the entire Gibbs space. See Fig. \ref{Fig1}. Under such highly peculiar circumstances, applicability of nonadditive entropic functionals is in the menu! For example, the infinity which, at the critical point, emerges for the Gruneisen parameter $\Gamma_{BG}$ disappears when calculated for an unique value of the entropic index $q^\star <1$ of the nonadditive entropic functional $S_q$, the value of $q$ such that  $S_{q^\star}$ is thermodynamically extensive (i.e., $S_{q^\star}(N) \propto N$ in the $N\to\infty$ limit).  More precisely, at the critical point, a $q$-generalized $\Gamma_q$ can be calculated which {\it diverges} for $q>q^\star$ (which includes the $q=1$ case, corresponding to the BG theory), it is {\it finite} for $q=q^\star$, and vanishes for $q < q^\star$. See Fig. \ref{Fig2}.
This special value $q^\star$  depends on the symmetry which is spontaneously being broken in that phase transition, and is therefore different for say the Ising, XY, Heisenberg ferromagnets. Consistently, within a given symmetry (say the Ising one), what changes instead is the value of $\Gamma_{q^\star}$; indeed, this value depends on nonuniversal variables such as the size of the spins (1/2, 3/2, or whatever), or on whether we have only nearest-neighbor coupling or say nearest- and next-nearest-neighbor coupling, and their respective intensities (see \cite{mariano1,mariano2}).

As a summarizing corollary, we may emphasize that using BG statistical mechanics at precisely the critical point indistinctively yields, for quantities such as the isothermal magnetic susceptibility, specific heat, correlation length, and others, a mere divergence, {\it for all models} with all kinds of spontaneously broken symmetries and all short-ranged couplings. This is to say, the BG theory makes no distinction at all about all these universal (symmetry, dimensionality $d$) and nonuniversal (diverse neighborhood couplings, size of microscopic spins, etc) ingredients. In sensible contrast, statistical mechanics grounded on nonadditive entropic functionals such as $S_q, S_\delta, \dots$  satisfactorily reflect these unambiguously important aspects of criticality. 

It is certainly worthy to quote here interesting (and definitively pertinent) statements about infinity by outstanding scientists and philosophers: 

{\it ... it is incumbent on the person 
who specializes in physics to discuss the 
infinite and to inquire whether there is such 
a thing or not, and, if there is, what it is.}\\
Aristotle [Physics, 350 BCE]\\

{\it Renormalization is just a stop-gap procedure.\footnote{A "stop-gap procedure" is a temporary solution or measure used to address 
an immediate problem until a permanent fix can be implemented;  
…quick and simple solution to keep a system or process working, but it is 
not intended to be permanent.
Its main purpose is to prevent a problem from getting worse or causing further damage while a better solution is sought.
} There must be some fundamental change in our 
ideas, probably a change just as fundamental as 
the passage from Bohr’s orbit theory to quantum 
mechanics. When you get a number turning out 
to be infinite which ought to be finite, you should 
admit that there is something wrong with your 
equations, and not hope that you can get a good 
theory just by doctoring up that number.}\\ Paul A. M. Dirac [circa 1950] \\

{\it The presence of infinities in physics always signals that we have missed some crucial point in our mathematical treatment.}\\
U. Harbach and S. Hossenfelder [Phys. Lett. B {\bf 632}, 379 (2006)]\\

This is the kind of question that I would have loved to calmly discuss with Hans... but, {\it h\'elas}, he left us too early!  My homage and affection to his memory!

\section*{Acknowledgments}
I warmly thank Ko van der Weele, 
Joseph Indekeu,
Andreas Schadschneider and the whole editorial team for kindly inviting me to contribute in the present Special Issue. I also thank E.M.F. Curado for an interesting discussion about the physical interpretation of the concept of entropy. 
Finally, I acknowledge partial financial support by the Brazilian Agencies CNPq and FAPERJ. 



\section*{Declaration of Competing interests}
The authors declare no competing interests.

\end{document}